\documentclass[preprint,journal]{vgtc}       
\ieeedoi{10.1109/TVCG.2019.2934431}

\ifpdf
  \pdfoutput=1\relax                   
  \usepackage{graphicx}                
  \DeclareGraphicsExtensions{.pdf,.png,.jpg,.jpeg} 
\else
  \usepackage{graphicx}                
  \DeclareGraphicsExtensions{.eps}     
\fi%

\graphicspath{{figures/}{pictures/}{images/}{./}} 
\usepackage{pdfpages}
\usepackage{microtype}                 
\PassOptionsToPackage{warn}{textcomp}  
\usepackage{textcomp}                  
\usepackage{mathptmx}                  
\usepackage{times}                     
\usepackage{cite}                      
\usepackage{tabu}                      
\usepackage{booktabs}                  
\usepackage{subfigure}

\usepackage{wrapfig}       
\usepackage{soul}       

\usepackage{listings}
\usepackage{color}
\definecolor{lightgray}{rgb}{.9,.9,.9}
\definecolor{darkgray}{rgb}{.4,.4,.4}
\definecolor{purple}{rgb}{0.65, 0.12, 0.82}
\definecolor{royalblue}{rgb}{0.25, 0.41, 0.88}

\lstdefinelanguage{JavaScript}{
	keywords={typeof, new, true, false, catch, function, return, null, catch, switch, var, if, in, while, do, else, case, break},
	keywordstyle=\color{blue}\bfseries,
	ndkeywords={class, export, boolean, throw, implements, import, this},
	ndkeywordstyle=\color{darkgray}\bfseries,
	identifierstyle=\color{black},
	sensitive=false,
	comment=[l]{//},
	morecomment=[s]{/*}{*/},
	commentstyle=\color{purple}\ttfamily,
	stringstyle=\color{royalblue}\ttfamily,
	morestring=[b]',
	morestring=[b]"
}

\onlineid{0}

\vgtccategory{Research}
\vgtcpapertype{system}

\title{Searching the Visual Style and Structure of D3 Visualizations}

 \author{Enamul Hoque and Maneesh Agrawala
}
 \authorfooter{
 \item
    Enamul Hoque is with York University. E-mail: enamulh@yorku.ca.
 \item
  Maneesh Agrawala is with Stanford University. E-mail:  maneesh@cs.stanford.edu.
 }

\usepackage{color}
\definecolor{darkgreen}{rgb}{0,0.5,0}
\definecolor{orange}{rgb}{1,0.5,0}
\definecolor{teal}{rgb}{0,0.5,0.5}
\definecolor{darkpurple}{rgb}{0.5, 0, 0.5}

\newcommand{\squishlist}{
 \begin{list}{$\bullet$}
  { \setlength{\itemsep}{0pt}
     \setlength{\parsep}{3pt}
     \setlength{\topsep}{3pt}
     \setlength{\partopsep}{0pt}
     \setlength{\leftmargin}{1.5em}
     \setlength{\labelwidth}{1em}
     \setlength{\labelsep}{0.5em} } }
\newcommand{\squishend}{
  \end{list}  }

\abstract{We present a search engine for D3 visualizations that
allows queries based on their visual style and
underlying structure.
To build the engine we crawl a collection of
7860 D3 visualizations from the
Web and deconstruct each one to recover its data, its
data-encoding marks and the encodings describing how
the data is mapped to visual attributes of the marks.
We also extract axes and other non-data-encoding attributes of marks
(e.g., typeface, background color).  Our search engine indexes this
style and structure information as well as metadata about the webpage
containing the chart. We show how visualization developers can search the collection to
find visualizations that exhibit specific design characteristics and
thereby explore the space of possible designs. We also demonstrate how
researchers can use the search engine to identify commonly used visual
design patterns
and we perform such a demographic design analysis across our
collection of D3 charts.  
A user study reveals that visualization developers found our style and structure based search engine to be significantly more useful and satisfying for finding different designs of D3 charts, than a baseline search engine that only allows keyword search over the webpage containing a chart.
} 


\CCScatlist{ 
 \CCScat{K.6.1}{Management of Computing and Information Systems}%
{Project and People Management}{Life Cycle};
 \CCScat{K.7.m}{The Computing Profession}{Miscellaneous}{Ethics}
}

\teaser{
   \vspace{-0.06in}
   \centering
   \includegraphics[width=1\linewidth]{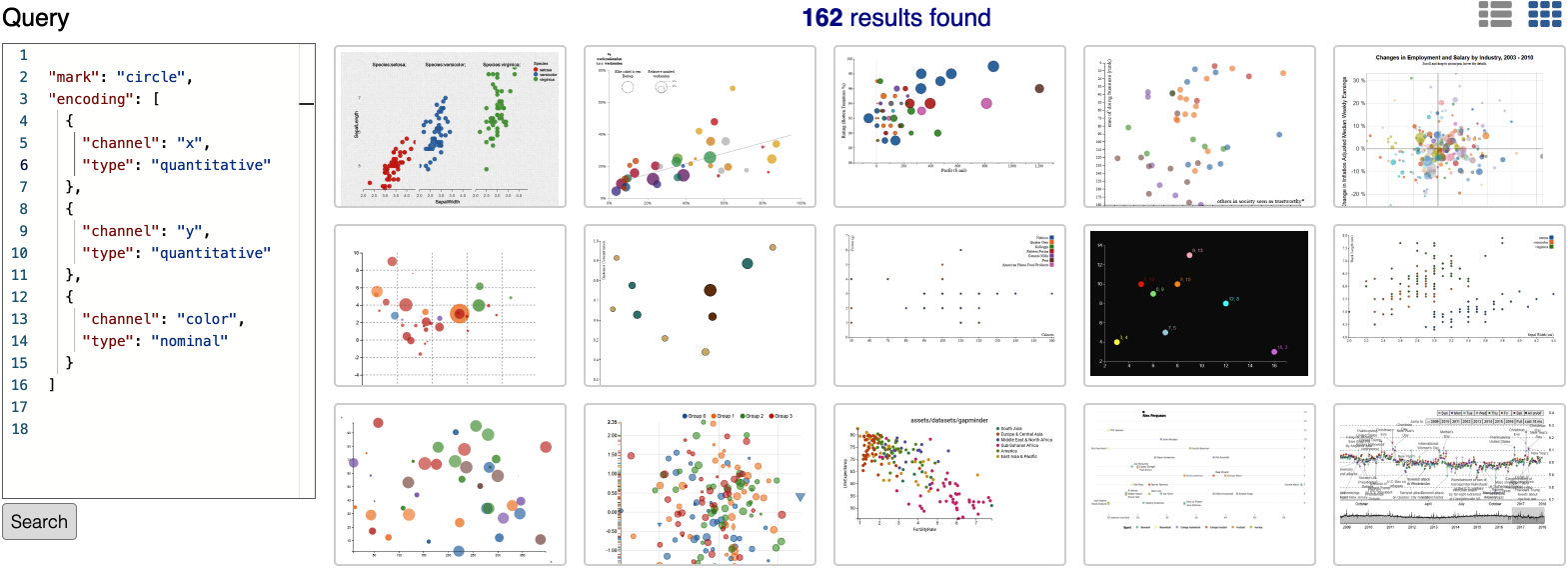}
      \vspace{-0.25in}
   \caption{ Our search interface lets users query for D3 
   	visualizations based on 
   	the data, the marks and the encodings that describe how visual attributes of the marks represent the data. Our query syntax is based on the Vega-Lite chart specification grammar~\cite{satyanarayan2017vega}. Here, the user searches for visualizations that use circles as data-encoding marks and contain at least three encodings. The query specifies that 
   	two of the encodings must map quantitative
   	data fields to the x-position and the y-position of the circle marks respectively,
        while a third encoding must map a
   	nominal data field to the fill-color of the
   	marks. The query results include a variety of
   	scatterplot and bubble chart styles that can help  developers explore
   	the design space of such visualizations.
   	\vspace{-.05in}
}
 	\label{fig:teaser}
 }

\vgtcinsertpkg

\begin{document}



\firstsection{Introduction}
\maketitle
The popularity of Web-based data visualizations has grown rapidly, as
people frequently use them to explore datasets and present complex
ideas.  Many of these visualizations are created using D3.js, a
JavaScript library\,\cite{d3js}, with a large community of developers.
Sites like the New York Times\,\cite{nytimes}, the Wall Street
Journal\,\cite{wsj}, Bloomberg\,\cite{bloomberg} and
Pudding.cool\,\cite{pudding} regularly publish D3 visualizations to
tell stories with data. Thousands of D3 charts are available online,
depicting many different datasets and exhibiting a wide variety of
visual designs.

\begin{figure*}[t!]
	\centering
	\includegraphics[width=1\linewidth]{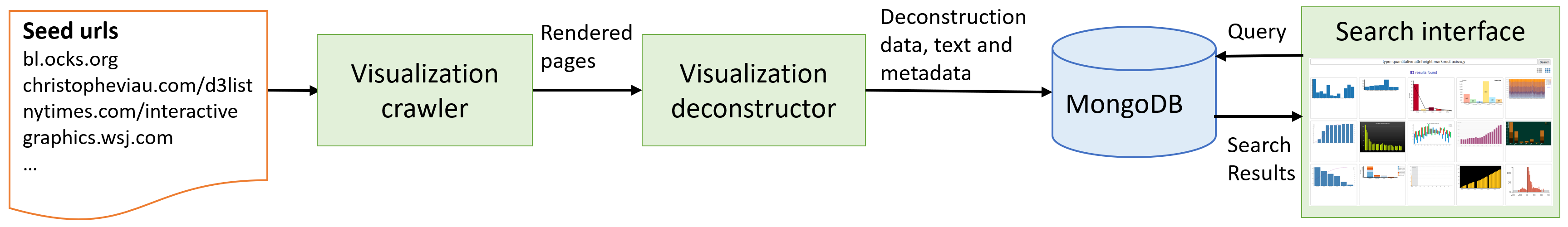}\\
                  \vspace{-0.15in}
	          \caption{Our visualization search engine is comprised of three main components (green boxes): a visualization crawler, a visualization deconstructor and a search interface. The crawler finds webpages containing D3 charts, the deconstructor extracts the visual style and structure of the charts (e.g. the data, the marks and the encodings) and the search interface lets users query for charts that include specific aspects of style and structure.}
                  \vspace{-0.18in}
	\label{fig:system}
\end{figure*}

The large collection of D3 visualizations available online serves as a
resource for visualization developers and researchers to
explore the space of visualization designs, and to analyze design
demographics.
For instance, visualization developers often seek example charts to
re-use or adapt the D3 code rather than writing the code from
scratch. Such developers also use online examples to better understand
design choices (e.g. encoding data using size vs. color) and stylistic
decisions (e.g. typefaces, color palettes).
Researchers may be interested
in analyzing the most frequently used design patterns (e.g. which
visual attribute -- position, length or color -- is most commonly used
to encode data?) or how often designers violate best practices
(e.g. limit the number of colors used to encode nominal data
to between six and twelve\,\cite{munzner2014visualization, Ware}). 
Such design patterns can provide a better understanding of
real-world design practices.

Today such design exploration and demographic analysis of Web-based D3
visualizations is difficult. Often the first step is to search for a
representative set of D3 charts.  But, generic search engines
(e.g. Google, Bing) index the entire Web and do not provide any way to
search only the pages containing D3 charts.  In contrast, the
Sightline\,\cite{sechler2017sightline} search engine and manually
curated D3 collections\,\cite{blocks, biglistd3} only include webpages
containing D3 charts.
While these tools make it easier for visualization developers
to find D3 charts, none of them provide access to the visual style,
structure and content of the charts. Therefore, it is impossible for
example to query for the set of D3 charts that encode data using
rectangle height, or to retrieve all of the charts that use a dark
background.  Instead developers and researchers must look at each D3
chart in the collection and manually check whether its visual design
satisfies the query.

In this paper, we present a search engine that indexes D3
visualizations based on their visual style and structure to directly
support such design queries. To build the engine we crawl a collection
of 7860 D3 charts from the Web and extend Harper and Agrawala's D3
Deconstructor\,\cite{harper2014deconstructing,harper2017converting}, to
recover their constituent data, data-encoding marks (e.g. rectangles
in bar chart, circles in a scatterplot) and the encodings describing
how the data is mapped to visual attributes of the marks (e.g. area,
height, fill-color). We also extract the visual attributes of axes
(e.g. attributes of tick marks and axis labels) and other
non-data-encoding attributes of marks (e.g. typeface, background
color) as these provide additional information about the style of the
visualization.  Our search engine indexes this style and structure
information as well as the text and metadata of the webpage containing
each chart.

Our query syntax is based on the
Vega-Lite\,\cite{satyanarayan2017vega}/CompassQL\,\cite{wongsuphasawat2017voyager} chart specification grammar
that declaratively represents a chart as a collection of encodings
that map data to visual attributes of marks. 
For example, as shown in
Figure\,\ref{fig:teaser}, users can search for all charts using circle
marks in which the x-position and y-position are mapped to
quantitative data fields while fill-color is mapped to a nominal data
field. This search yields a set of scatterplots and bubble charts that
contain at least these three encodings.  We show how such search
results can help visualization developers explore the variety of
charts that can be produced for a given set of encodings or design
characteristics.
We also demonstrate how researchers can use our search engine to
identify common visual design patterns and we perform such a
demographic design analysis across our collection of charts.
A user study comparing our search interface which indexes the style
and structure of D3 charts, with a baseline interface
(Sightline\,\cite{sechler2017sightline}) that only provides keyword
search over the webpage containing a D3 chart, reveals that
visualization developers find our interface significantly more useful
and satisfying to use as it allows them to quickly find different
designs and styles of D3 charts.

\section{Related Work}
{\bf Deconstructing visualizations:}
Constructing a chart involves encoding the data by mapping it to some visual
attributes (e.g. position, color) of 
marks (e.g. circles, rectangles). Deconstructing charts is the inverse process -- i.e., given a
chart, the goal is to extract the underlying data, the marks
with their visual attributes, and mappings (or encodings) between them.
There has
been growing interest in automatically deconstructing 
visualizations from bitmap images of charts.
ReVision\,\cite{savva2011revision} identifies chart type using an
SVM-based classifier then extracts the underlying
marks and data from an input chart image.
ChartSense\,\cite{jung2017chartsense} 
uses a convolutional neural network (CNN) for chart type classification
and a mixed-initiative approach to extract the
marks and data. Others have focused on 
extracting text from chart images (e.g., axis
labels)\,\cite{poco2017reverse}, recovering color encodings by
by analyzing the legend of a
chart\,\cite{poco2018extracting} and recovering data from scatterplots\,\cite{cliche2017scatteract}.
These techniques are orthogonal to
our work as they are designed to recover data, marks and encodings from
bitmaps of charts, while our goal is to index visualizations based
on the deconstructed data, marks and encodings.
%
As these methods improve in accuracy we believe it should be possible
to use them in conjunction with our methods to index the visual style
and structure of bitmap charts.




More recently, Harper and
Agrawala\,\cite{harper2014deconstructing,harper2017converting} have
developed techniques for deconstructing D3 visualizations.  Because
they work from a programmatic description of the chart they can
extract the data, the marks and the encodings far more accurately than
bitmap-based techniques.
Our work extends their deconstruction approach to extract additional
features including non data-encoding marks and their visual
attributes. Our search engine indexes this information and our query
syntax enables search over a broad range of the chart design space,
covering both style and structure.

\vspace{0.05in}
\noindent
{\bf Indexing and searching visualizations:}
Several researchers have developed techniques for indexing,
and searching Web-based visualizations. For example,
Sightline\,\cite{sechler2017sightline} is a chrome extension that
automatically indexes data visualizations encountered by users.
It
then supports retrieval based on keyword search over the text and metadata
(e.g. author, title) of the
webpage containing the visualization.
Graphical Histories\,\cite{heer2008} and
  KnowledgePearls\,\cite{2018_vast_knowledge-pearls} are visual
  analysis tools that record user interactions and the state of the
  visualization after each step of the analysis in a provenance
  graph. Users can then query for previous exploration states.
Others have developed techniques for searching for charts
in a collection of documents based on their text labels (e.g., axes
labels, titles, legends, and captions)\,\cite{diagramFlyer-chen-15,siegel2016figureseer,poco2017reverse}.
Lee et al.\,\cite{lee2016viziometrix} extract images from a large
corpus of scientific papers and classify the figure types (e.g.,
equation, diagram, plot and table).
Unlike these techniques our search engine
indexes lower-level aspects of visual style and structure, allowing
search by data, mark type and encoding.




\vspace{0.05in}
\noindent
{\bf Analyzing design demographics:}
Our work is inspired by recent efforts on mining large corpora of Web
documents to facilitate demographic analysis in various domains such
as webpage design\,\cite{kumar2013webzeitgeist,ritchie2011d} and API
usage\,\cite{glassman2018visualizing}. However, analysis of
visualization design demographics has been
rare. Beagle\,\cite{Begel-Baettle-2017}
classifies
several thousands SVG-based data visualizations 
by chart type (e.g. bar
chart, scatterplot). They find that the majority of charts
in the collection only cover four types: bar charts,
line charts, scatterplots, and geographic maps. We conduct a broader
demographic analysis of visualization design covering the usage of
different mark types and encodings.

\begin{figure*}[t]
  \centering
  \vspace{-0.05in}
  \includegraphics[width=1\linewidth]{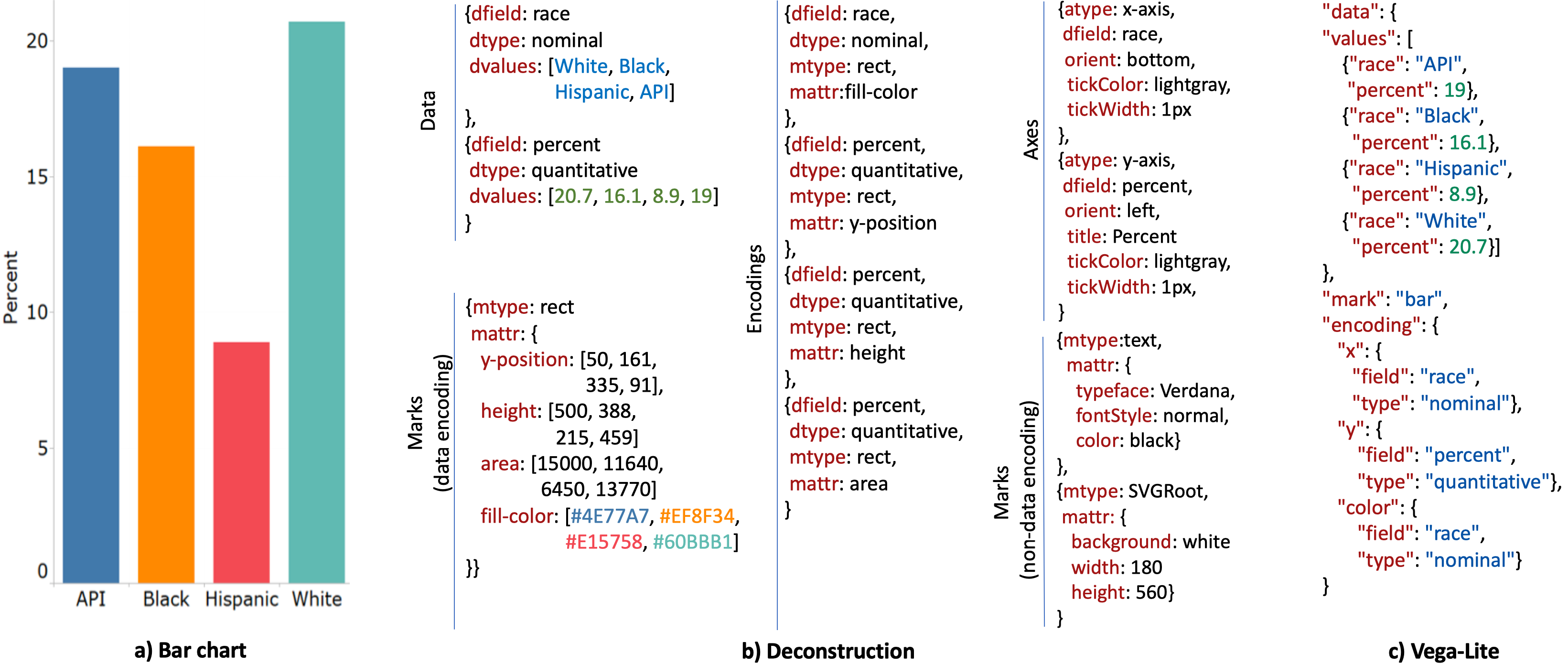}\\
    \vspace{-0.1in}
	\caption{A D3 bar chart (a). The data, marks, encodings, axes and non-data-encoding attributes of marks as extracted by our deconstructor (b). In this case, our deconstructor extracts two data fields named ``race'' and ``percent'' of data types ``nominal'' and ``quantitative'' respectively. It also extracts data-encoding marks of type ``rect'' and a set of 4 encodings that map data fields to visual attributes of the rect marks (e.g., the data field ``race'' is mapped to the fill-color of ``rect''). Finally, it extracts the axes, including the data field they represent and their orientation and non-data-encoding attributes of the marks (e.g. typeface, background color).  A Vega-Lite specification for the bar chart is similar to our deconstructed representation and explicitly describes the relevant encodings (c). 
	}
        \vspace{-0.2in}
	\label{fig:vis_example}
\end{figure*}

\vspace{-.02in}
\section{Search Engine for D3 Visualization}

As shown in Figure~\ref{fig:system}, our search engine consists of
three major components: (1) The {\em visualization crawler} downloads
webpages starting from a collection of seed URLs. (2) The {\em
  visualization deconstructor} extracts the visual style and structure
of the D3 visualizations found on each crawled webpage and stores the
deconstructed information in a database. (3) The {\em search
  interface} then lets end-users query the database based on the style
and structure information.





\vspace{-.07in}
\subsection{Visualization Crawler}

Our visualization crawler extracts weblinks from a set of seed URLs, then downloads the corresponding webpage and finally renders the resulting page in the Chrome browser so that 
we can apply our visualization
deconstructor to the page. Our crawler only follows one level of
links, and we seed it with webpages containing lists of links to D3
visualizations. Members of the D3 developer community have compiled
several such lists at sites like d3.js~\cite{d3js}, 
bl.ocks.org~\cite{blocks} and the ``Big List of D3
Examples''~\cite{biglistd3}.  We supplement these community curated
seed lists with additional lists of interactive visualizations that
are available directly from news sites like the New York
Times~\cite{nytseed} and the Wall Street Journal~\cite{wsj}.

In total we crawled 2623 webpages from 622 different domains, where
each page contained at least one D3 visualization, and we
deconstructed a total of 7860 D3 visualizations.
We also captured screenshots of the
visualization as well as the HTML of the webpage it was embedded
in. The resulting database takes 5.6 GB of disk storage. 
Supplemental Materials: Figure 1 shows the top 20 domains in terms of the number of D3 visualizations we extracted from them.

\subsection{Visualization Deconstructor}
\label{sec:deconstruction}

The goal of our visualization deconstructor is to extract the complete
visual style and structure of every D3 chart that appears on an input
webpage (Figure~\ref{fig:vis_example}).
%
Each data field in our deconstruction is 
is represented by a field name \texttt{``dfield''}, data
type \texttt{``dtype''}, and a set of data values \texttt{``dvalues''}.
In Figure~\ref{fig:vis_example}, one
data field is named {\em race} with data type {\em nominal}, while the
other data field is named {\em percent} with data type 
{\em quantitative}.  The data encoding marks are represented by their mark
types \texttt{``mtype''} which can take values like {\em rect, circle,
  line, path, and text} as well as mark attributes \texttt{``mattr''} which can take values like {\em fill-color, opacity, size, angle, x-position} and {\em y-position}. Each data-encoding mark attribute can also take a set of values. For example {\em y-position} has four values each of which corresponds to the  {\em y-position} of a {\em rect} mark in screen space.
Our deconstructor
analyzes the relationships between data values and the values of each
mark attribute to recover the encodings that map between them -- represented as
\texttt{\{dfield, dtype, mtype, mattr\}}. For example,
the encoding at the top of Figure~\ref{fig:vis_example}b (middle column) specifies that 
the {\em race} data field is mapped to the {\em fill-color} of the data-encoding {\em rect} marks.

Our deconstructor also extracts each axis, including the axis type
\texttt{``atype''} either {\em x-axis} or {\em y-axis}, the data field
\texttt{``dfield''} it represents, its orientation \texttt{``orient''}
either {\em left, right, bottom} or {\em top} and other axis
attributes such as \texttt{``title''}, \texttt{``tickColor''},
\texttt{``tickWidth''}, etc.  Finally it extracts non-data-encoding
attributes of the marks such as {\em typeface} and {\em background}
color.  Note that our deconstructor treats the root SVG node of the D3
chart a non-data-encoding mark and extracts its attributes as well.



Our deconstructor builds on Harper and Agrawala's D3
Deconstructor~\cite{harper2014deconstructing,harper2017converting}
which focused on recovering the data, the marks and the encodings from
a D3 chart. We extend their deconstructor in three key ways: (1) We
extract an additional visual attribute \textit{angle} for marks
instantiated using D3's \texttt{d3.arc()}. Such {\em arcs} are
typically used to create pie charts, donut charts and radial bar
charts which encode data using the {\em angle} of the {\em arc}.
(2) We extract visual
attributes that do not typically encode data but are used instead for styling (e.g.,
background color, typeface, stroke-width) from all the
marks and SVG elements including text elements and the top-level SVG
node. (3) We extract axes and their attributes such as the data fields they represent, their orientation, their title text, etc.
The resulting deconstruction is very similar to a Vega-Lite specification which
focuses on the encodings between data and mark attributes (Figure~\ref{fig:vis_example}c)
We describe these extensions 
in detail
 in the supplemental material. 
We store the 
   deconstruction data
	 in a
  MongoDB database~\cite{mongodb}. We chose MongoDB since it is suitable for processing the  hierarchical structure in JSON format that our deconstructor generates.


\subsection{Search User Interface}
\label{sec:interface}
Our search interface lets users query the database of deconstructed D3
visualizations by providing a partial specification of the desired
results. We first describe our query syntax and then present the
interface features for exploring the search results.

\begin{figure*}[t]
	\centering
        \vspace{-0.1in}
	\includegraphics[width=1\linewidth]{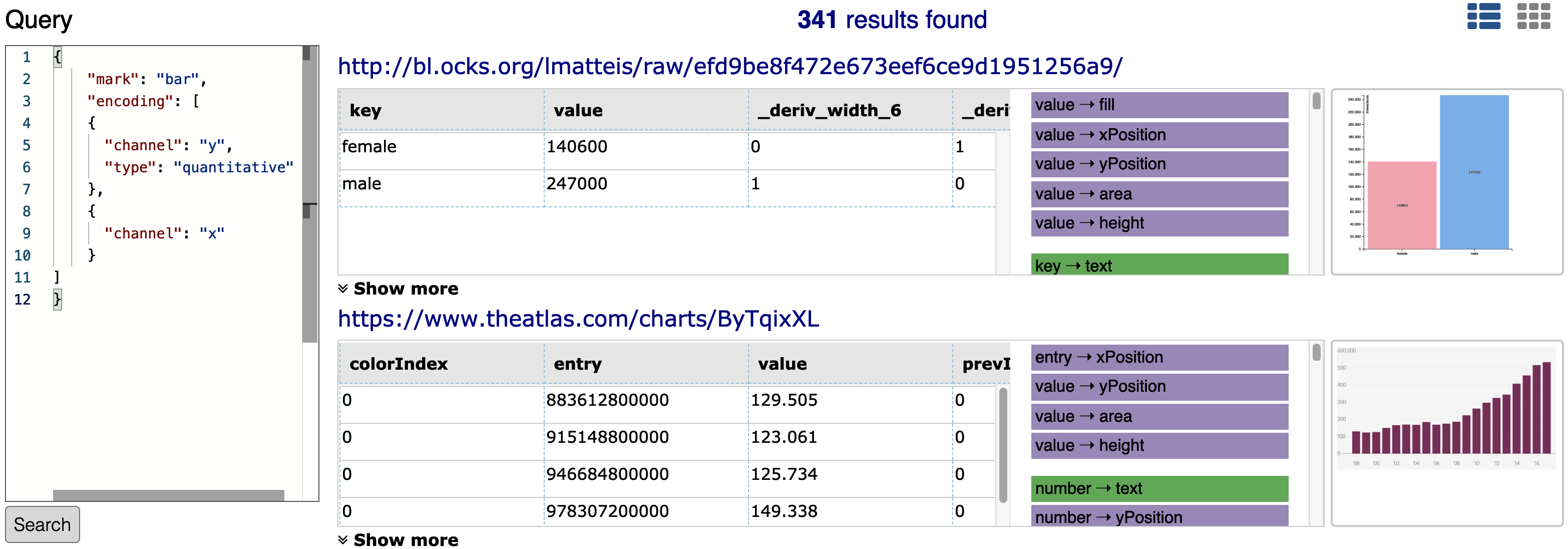}\\
        \vspace{-0.05in}	
	\caption{Query for charts containing one quantitative data field using the y-position and another data field of either quantitative or nominal data type using the x-position of the bars. The search results are presented in list view showing the data table and encodings between data fields and mark attributes.
          The URL for the webpage containing the chart is given above the data table and a thumbnail shows the visual form of the chart and this query returns a set of vertical bar charts.
          As in Harper and Agrawala~\cite{harper2014deconstructing,harper2017converting} each encoding is represented as \textit{dfield} $\rightarrow$ \textit{mattr} and the background color indicates the data type: green for nominal data and purple for quantitative data.  
 Users can click on `Show more' button to expand the view so that all the data tables and encodings are visible.
        }
        \vspace{-0.2in}
	\label{fig:search_interface}
\end{figure*}

\subsubsection{Query syntax}
Our query syntax is based on the Vega-Lite~\cite{satyanarayan2017vega}
grammar for declaratively specifying a chart as a collection of
encodings that map data to visual attributes of marks
(Figure~\ref{fig:vis_example}c). We chose the Vega-Lite syntax because
it explicitly describes the structure and style of a visualization and
is similar to other grammar-based representations~\cite{ggplot,
  wilkinson2012grammar}.  Grammar-based specifications like Vega-Lite
have proven to be an effective representation for conceptualizing
visualizations~\cite{wilkinson2012grammar} because they help
developers understand the way charts are constructed from data, marks
and encoding.  Moreover, Vega-Lite is rapidly becoming a popular tool
for creating charts as it has been incorporated into Jupyter
Notebooks~\cite{jupyter} and Python more generally through
Altair~\cite{altair}.  Thus, developers who are familiar
with Vega-Lite can form search queries without having to learn a new
syntax and for those unfamiliar with Vega-Lite, our interface can help
them learn a useful representation.

Note however, that unlike Vega-Lite which requires users to provide a
complete specification to produce a chart, our query syntax allows
partial specification. Queries may include only one or more aspects of
the data, marks, encodings, axes, and non-data-encoding attributes of
marks.  Our query syntax also includes functional operators
(e.g. \texttt{count, gt, lt}) and logical operators (e.g. \texttt{and,
  or, not}) on values and supports
CompassQL's~\cite{wongsuphasawat2017voyager} syntax for replacing
concrete values with wildcards. 




 \vspace{0.05in}
\noindent
\textbf{Data:} Queries about data can be based on the name of the data
field or on data values. 
For
example, users can search for visualizations the include a data field
name ``population'' with the query
 \vspace{-0.05in}
\begin{verbatim}
{"data":{"field": "population"}}
\end{verbatim}
 \vspace{-0.05in}
Although Vega-Lite does not include the \texttt{"field"} key as part
of the \texttt{"data"} specification,  it does include \texttt{"field"} as part
of an \texttt{"encoding"} specification
(Figure~\ref{fig:vis_example}c). We allow \texttt{"field"} as part of
a \texttt{"data"} specification to enable queries based on field names that may
be present in the dataset but not encoded in the corresponding
visualization. Also note that our syntax allows any concrete string to
be specified as a regular expression. For instance, we could replace
``population'' with ``.*population.*'' to find all visualizations that
include a data field with the word ``population'' in the field name.

Our query syntax also allows users to apply functional operators on
data values including \texttt{``count''} which computes the number of
data values, statistical calculations on quantitative values
\texttt{``min''}, \texttt{``max''}, \texttt{``sum''} and \texttt{``average''} and
comparisons (\texttt{``gt''}, \texttt{``gte''}, \texttt{``lt''}, \texttt{``lte''}).
For example, the following query
 \vspace{-0.05in}
\begin{verbatim}
{"data":{"values":{"count": {"gte":1000}}}}
\end{verbatim}
 \vspace{-0.05in}
returns visualizations containing at least 1000 data points. 
Finally, our syntax allows users to apply logical operators on an array of values. For example, the query
 \vspace{-0.05in}
\begin{verbatim}
{"data":{"field":"country", 
         "values": {"and": ["India","China"]}}}
\end{verbatim}
 \vspace{-0.05in}
finds visualizations that have a data field named ``country'' with values including both ``India'' and ``China''.


\vspace{0.05in}
\noindent
\textbf{Marks:} Queries about marks can be based on the mark type. For example,
the query \texttt{\{"mark": "circle"\}}
would find all visualizations that include marks of type circle.  Note
that while D3 does not explicitly include the mark type {\em bar},
Vega-Lite does include it to enable concise specification of bar
charts. Like Vega-Lite, we also consider {\em bar} as a mark
type. Specifically after deconstruction, we convert marks of type {\em
  rect} that encode quantitative data using length and maintain a
fixed width, to marks of type {\em bar}.

\begin{figure*}[t]
  \centering
  \vspace{-0.092in}
	\includegraphics[width=1\linewidth]{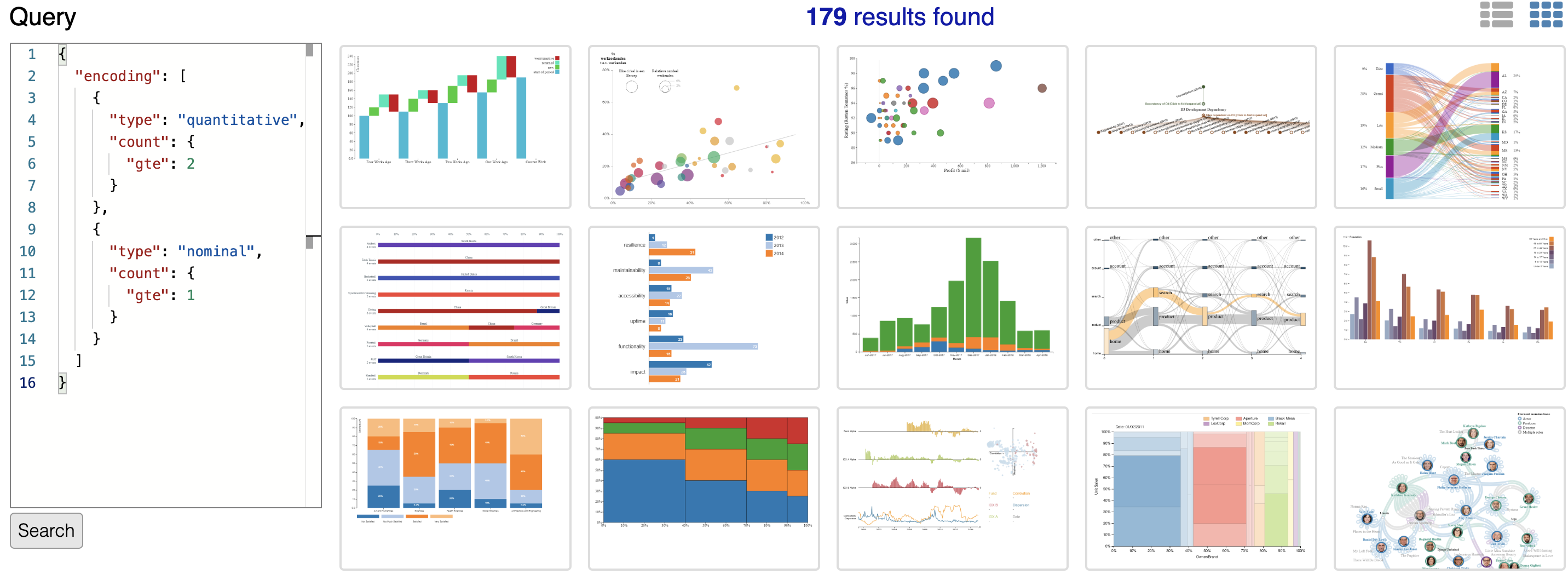}\\
  \vspace{-0.1in}	
	\caption{Query for charts containing at least one encoding of nominal data fields and at least two encodings of quantitative data fields. Our search engine returns stacked and grouped bar charts, bubble charts and other variations with the desired encodings.}
        \vspace{-0.2in}
	\label{fig:2quant1cat}
\end{figure*}

\vspace{0.05in}
\noindent
\textbf{Encodings:} Queries about encodings can be based on the
data-encoding mark attribute (called \texttt{``channel''} in CompassQL~\cite{wongsuphasawat2017voyager}),
the data type or the data field name.
Queries specifying a \texttt{``channel''} can search for encodings using 
any data-encoding mark attribute value including {\em color} (i.e. fill-color), {\em opacity}, {\em shape}, {\em text}, {\em angle}, {\em size} (i.e. area), {\em x}-position or {\em y}-position. 
With this syntax, the query
 \vspace{-0.08in}
\begin{verbatim}
{"mark": "bar",
 "encoding": [
    {"channel":"y","type":"quantitative","field":"*"}, 
    {"channel": "x","type": "*","field": "*"} ]}
\end{verbatim}    	
 \vspace{-0.08in}
finds all charts that encode a quantitative data field using the
y-position of bar marks and encode another data field using the
x-position of the bars.  As in CompassQL, the \texttt{``*''} symbol is a wildcard
indicating that the data field encoded using y-position can have any
data field name. Similarly, the data field encoded using x-position can
have any data field name and be of any data type.
Note that to reduce the verbosity of our syntax, if the user does not include a complete Vega-Lite specification we treat the unspecified part of the query as a wildcard by default. Thus we could equivalently write this query as
 \vspace{-0.05in}
\begin{verbatim}
{"mark": "bar",
 "encoding": [
    {"channel": "y", "type": "quantitative"}, 
    {"channel": "x"} ]}
\end{verbatim}
 \vspace{-0.05in}
As shown in Figure~\ref{fig:search_interface} this query retrieves vertical bar charts.

In addition to wildcards, our query syntax supports pattern
variables. These variables start with the symbol \texttt{\$} and can be used
in multiple places to indicate that the search engine should bind the same value in each place it appears. 
For example, the query
 \vspace{-0.05in}
\begin{verbatim}
{"encoding": [
    {"channel":"x","field":"$t","type":"quantitative"}, 
    {"channel":"color","field":"$t","type":"*"}]}
\end{verbatim}    	
 \vspace{-0.05in}
finds visualizations in which the same data field is redundantly
encoded using the x-position and fill color of a mark. If we replace the
pattern variable \texttt{``\$t''} with a wildcard \texttt{``*''} the query would return
visualizations in which either the same or different data fields are
encoded using x-position and color respectively.


\vspace{0.05in}
\noindent
\textbf{Axes:} In Vega-Lite an encoding may also be used to create and
specify properties of axes.
For example 
the syntax
 \vspace{-0.05in}
\begin{verbatim}
{"encoding":[{"channel":"x", "axis": true}]}  
\end{verbatim}
 \vspace{-0.05in}
can be used in our search engine to find visualizations that contain an x-axis. Users can similarly
add axis properties to the query. For instance, adding \texttt{"orient":
  "bottom"} to the axis encoding would find charts containing
an x-axis near the bottom of the visualization.



\vspace{0.05in}
\noindent
\textbf{Non-data-encoding attributes:}
To search for charts containing specific non-data-encoding attributes
users can
provide queries like
 \vspace{-0.05in}
\begin{verbatim}
{"mark": {"type": "text", "typeface": "Verdana"}}
\end{verbatim}
 \vspace{-0.05in}
In this case, the query finds visualizations that contain at least one
mark of type {\em text} that uses the \texttt{``typeface''} {\em Verdana}.


\vspace{0.05in}
\noindent
\textbf{Metadata:} In addition to the properties of visualization 
users can search the metadata of the webpages that contain the 
visualizations.  Our query syntax supports 
keyword search on the surrounding webpage text
(using \texttt{"keyword''}), webpage \texttt{``title''}, \texttt{``url''}, \texttt{``domain''} name, and the \texttt{``timestamp''} of
the webpage if it contains one in its header. For example, the query

 \vspace{-0.08in}
 \begin{verbatim}
{"domain": ".*nytimes.*"}}
\end{verbatim}
 \vspace{-0.08in}
finds all visualizations, where domain contains ``nytimes''.

%
%
%

 \subsubsection{Search Engine Interface}
 The interface to our search engine includes a query textbox and
 either a grid of thumbnails view (Figure~\ref{fig:teaser}) or list
 view (Figure~\ref{fig:search_interface}) of results.
 The interface initially shows the query in Figure~\ref{fig:search_interface}
 as an example of our query syntax that users can modify to formulate their own query. This design helps users who are learning the Vega-Lite syntax by supporting recognition of the syntax rather than having to recall it.
To further help users quickly formulate queries our query textbox provides
automatic code-formatting, auto-completion and validation using Visual
Studio Code~\cite{vs-code}, an open source code editor library we
access via a JavaScript wrapper.  Visual Studio Code allows us to
associate a schema describing the allowable values for each key in our
query syntax (e.g. ``channel", ``field'', ``type'').  Our interface then
triggers auto-complete suggestions as the user is typing based on this
query syntax scheme. It also validates the query by checking the
structure and values in the query with respect to the schema and then
highlights potential errors.  

Our search interface can display the search results using two
different ordering strategies.
The {\em ranking} strategy counts the number of encodings in the
search results that match the encodings explicitly specified in the query.
We consider a pair of encodings
to match if they both map data of the same data type to the same
visual attribute of the same mark type.
We rank the search results based on the number of matched
encodings. If there is a tie in the number of matched encodings we
consider how many encodings were unmatched, where more unmatched
encodings indicates less relevance. Since search results generally
match all of the explicit encodings in the query, this approach
primarily relies on the tie-breaking approach to produce the final
ordering.
The {\em randomized} strategy randomly orders the search results and
is designed to show users a variety of different designs that match
the query. We use the randomized strategy by default.




\begin{figure*}[t!]
	\centering
	\includegraphics[width=1\linewidth]{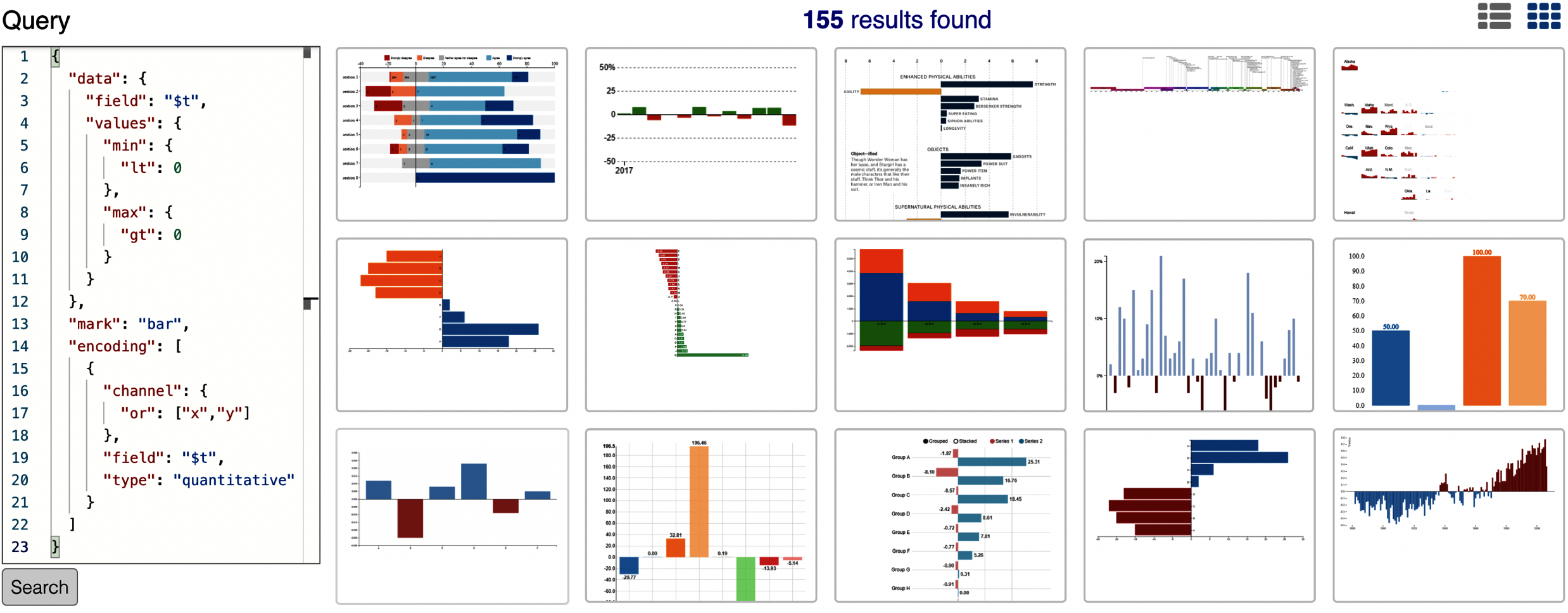}\\
	 \vspace{-0.1in}
	\caption{Query for charts containing
		a quantitative data field with both negative and positive data values mapped to the x- or y-position attribute of bar marks. Our search engine mostly returns bar charts with bars extending in both directions from a baseline at zero. 
		The user could filter out this result by for example including another condition in the query checking for axes.}
	\vspace{-0.12in}
	\label{fig:bar-negative}
\end{figure*}

\section{Applications}
Our search engine currently supports two main classes of applications:
(1) design space exploration, and (2) information seeking.

\subsection{Design space exploration}
Often there are many ways to visualize the same dataset. In order to
find a representation that is effective and expressive, visualization developers
typically explore a variety of design choices. Our search engine
supports developers in exploring the design space by enabling a range
of queries based on visual style and structure of D3 charts.

\vspace{0.05in}
\noindent \textbf{ Data, marks, and encodings:}
Given a dataset with a particular combination of data types,
visualization developers often need to consider the possible
alternatives for depicting the data. For example, suppose a dataset
contains one nominal and two quantitative data fields. A developer
might issue the query shown in Figure~\ref{fig:2quant1cat}, 
to find charts that encode at least one nominal and two quantitative
data fields.  In this query the
\texttt{count} and \texttt{gte} operators are only necessary when
querying for more than one encoding of the same type. A query for
\texttt{"encoding": [\{"type": "nominal"\}]} would return charts that
encode at least one nominal data field.

Users can search for different basic chart types (e.g., bar charts,
scatterplots) by querying for combinations of encodings.  For example,
to retrieve bar charts, the query in Figure \ref{fig:search_interface}
looks for charts that encode quantitative data using the y-position
attribute of bars and contain both x- and y-axes.  Similarly, the
query in Figure~\ref{fig:teaser} looks for scatterplots that encode an
additional data fields using the color of circle marks.

Users can also query based on the values of the data fields.  Consider
a visualization developer who has a quantitative data field with both
positive and negative values and wants to create a bar chart with a
baseline at zero, such that the positive bars extend upwards (or to
the right for a horizontal bar chart) and negative bars extend
downwards (or to the left).  A novice developer may find it tricky to
implement such a bar chart because rectangles in D3 are positioned
using their top-left corner and cannot have negative heights or
widths.  Using our search engine the developer can query to find a
variety of bar charts where both positive and negative values are
encoded as in Figure~\ref{fig:bar-negative}.
The query looks for charts that map a quantitative data field to
either the x- or y-position of bars.
The min and max
conditions indicate that the data field depicted in the chart
should have both negative and positive values.  The pattern variable
\texttt{``\$t''} is used
to ensure that the same data field containing negative and
positive values is encoded using the x- or y-position
attributes. The
resulting bar charts include a number of stylistic
variations of bar charts with negative and positive values.

\begin{figure*}[t!]
	\centering
	\includegraphics[width=1\linewidth]{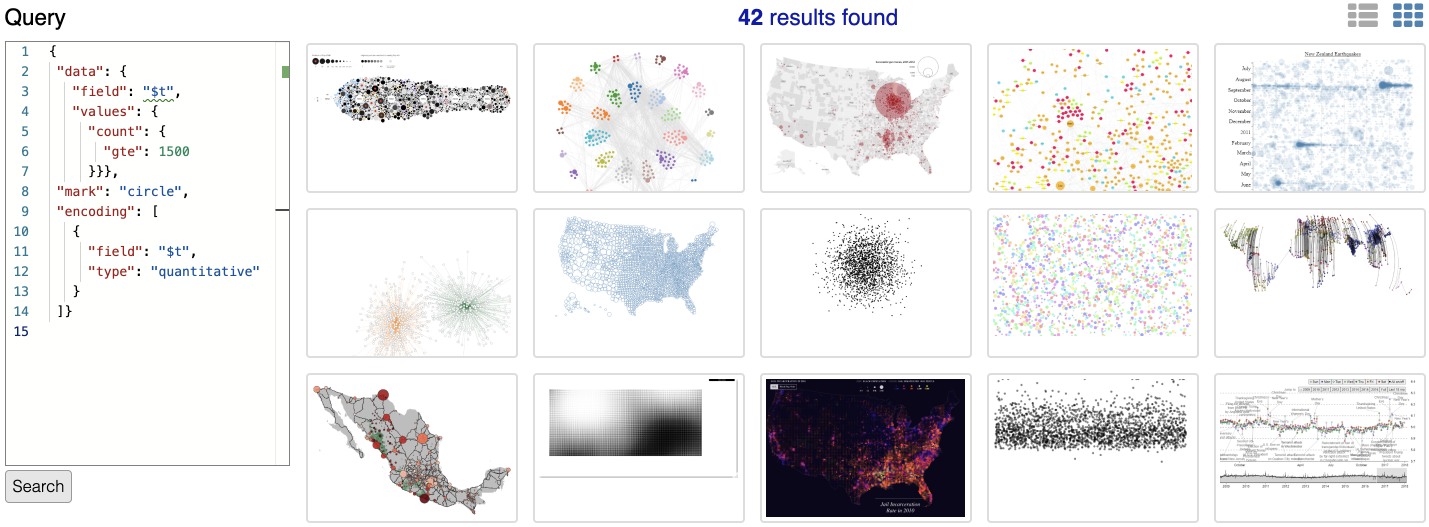}\\
	\vspace{-0.07in}
	\caption{Query for visualizations that contain at least 1500 data points of a quantitative data field encoded using some visual attribute of circle marks. The results include scatterplots, bubble charts and other more exotic chart types.}
	\vspace{-0.2in}
	\label{fig:dense_vis}
\end{figure*}

Developers sometimes face the problem of visualizing large datasets
containing thousands of data points in limited screen space.  In such
cases, It can be useful to see a gallery of example visualizations to
see how other visualization developers dealt with similarly large datasets. With
the query shown in Figure~\ref{fig:dense_vis},
our search engine returns charts in which at least one quantitative
data field contains at least 1500 data points encoded using circle
marks. Since the \texttt{"channel"} is
not specified in the query, any possible visual attribute
(e.g. positions, area, and color) may be used to encode the data.

\begin{figure*}[t!]
	\centering
	\vspace{-0.09in}
	\includegraphics[width=1\linewidth]{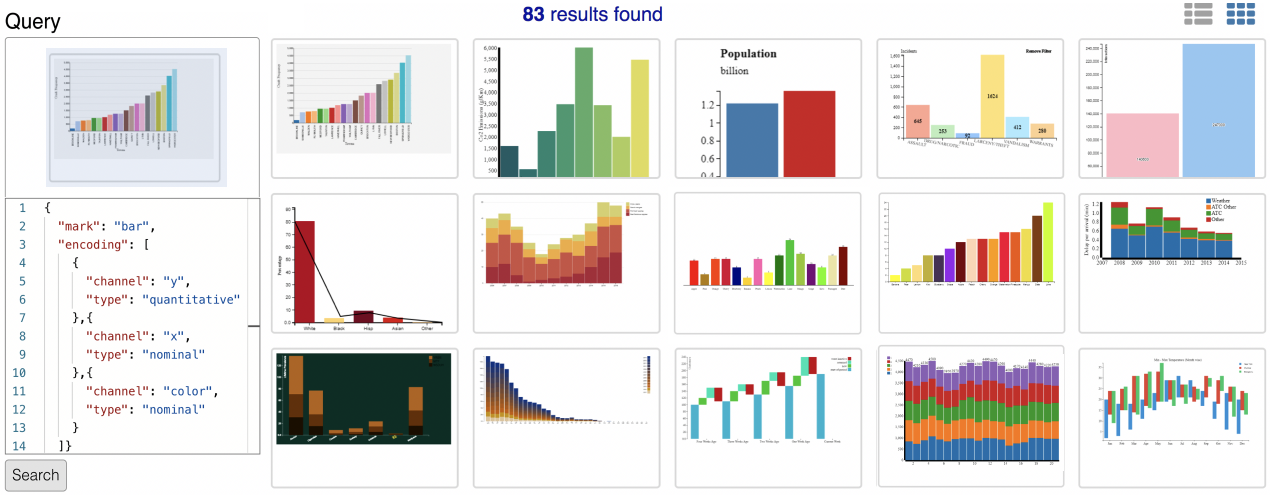}\\
	
	\vspace{-0.13in}
	\caption{In query by example users select a chart
		returned by an earlier query (shown at top of query textbox) and our search engine finds charts with similar encodings. Here, the example query is a vertical bar chart that contains an additional encoding that maps a nominal data field to the fill-color of the bars. Our search engine automatically converts the example into our query syntax keeping only the encodings (shown at bottom of query textbox). The query returns a variety of vertical bar charts that match as many of the encodings as possible. }
	\vspace{-0.12in}
	\label{fig:query_by_vis}
\end{figure*}

\vspace{0.05in}
\noindent \textbf{Query by example:} Developers sometimes seek
visualizations that are similar to an example visualization.  For
example, suppose a visualization developer is interested in seeing the stylistic
variations amongst bar charts that use the fill-color of the bars to
depict an additional nominal data field. Alternatively, the developer
may wish to examine different ways to implement this type of bar
chart.  Our search engine lets users select an example visualization
-- e.g. a bar chart with an additional fill-color encoding --
and returns similar visualizations based on the similarity of their
marks and encodings as in Figure~\ref{fig:query_by_vis}.

To implement this functionality, when users provide an example chart
as the query, our system deconstructs it
(Section~\ref{sec:deconstruction}) and then forms a query consisting
of the mark and encoding specifications found in the example.
However, the query omits data field names, axis properties and
non-data encoding mark properties to allow variations in the results
set.  Thus, users can see stylistic variations (e.g., different
background colors, typefaces or usage of grid-lines) in the thumbnail
view and go to the source webpage to examine the underlying code.

\begin{figure*}[t!]
	\centering
	\includegraphics[width=1\linewidth]{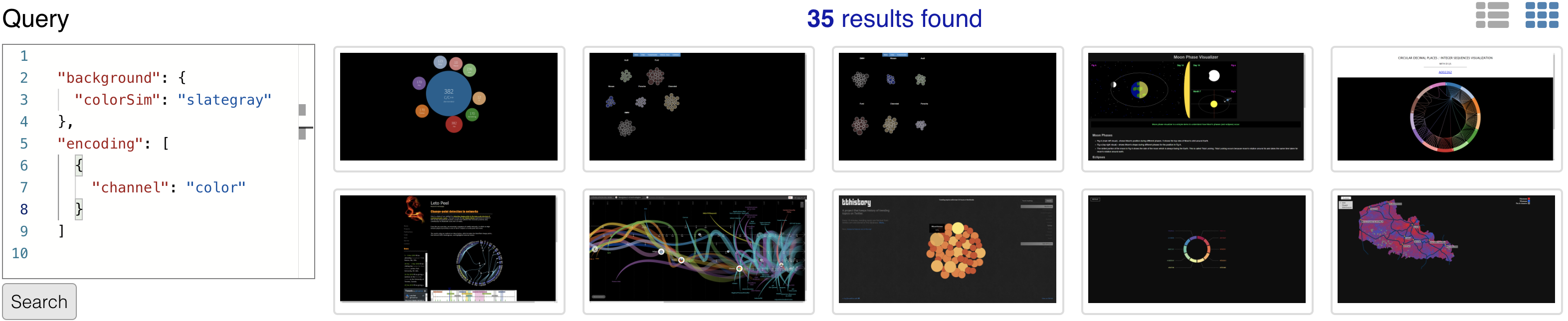}\\
	\vspace{-0.118in}
	\caption{ Query for charts containing dark backgrounds similar in color to ``slategray'' and that encode data using the fill-color of foreground marks. The results include a variety of color palettes for the foreground marks.}
	\label{fig:color_query_darkslategray}
	\vspace{-0.22in}
\end{figure*}

\vspace{0.05in}
\noindent \textbf{Query by non-data-encoding attributes}:
Charts often include visual attributes
that do not encode data but are used for styling such as the background color,
typefaces, stroke-widths, etc.
Consider for example, a developer who wants to create a chart
with dark background and needs to find suitable colors for filling the
foreground data-encoding marks.
The query in 
Figure~\ref{fig:color_query_darkslategray}
finds charts where the background color is similar to the named color
{\em ``slategray''} and that contain foreground data-encoding marks
that encode data using fill-color.  We use the functional operator
\texttt{"colorSim"} to compute the color similarity as the distance
between the background color of each chart in our database and {\em
  ``slategray''} in LAB color space.  When using this operator our
search engine rank orders the resulting charts based on this distance
and the developer can see how the color choices affect the visual
design.

\subsection{Information seeking}
Sometimes users wish to find visualizations that are related to a
specific topic. For example, a data journalist may seek charts
about the latest US election. While keyword-based search engines like
Google or Sightline~\cite{sechler2017sightline} can return some
visualizations on such topics, our search engine further enhances the
information seeking process by complementing text-based search
with retrieval based on mark and data attributes.  For example, the query
 \vspace{-0.05in}
\begin{verbatim}
{"keyword": "US election",
 "encoding": [{
     "channel": "color", "type": "nominal",
     "values": {"and":["red", "blue"]} }]}
\end{verbatim}
 \vspace{-0.05in}
finds charts that appear on webpages containing the word ``US election''
and that use blue and red fill-colors to encode nominal data
(Supplemental Materials: Figure 2).

Our search engine also lets users find visualizations based on data
field names and titles. However, sometimes a data field may not have
the exact same name that the user specified and instead use a
semantically similar synonym. For such cases we provide the
\texttt{"wordSim"} operator which computes semantic similarity to any
given word. Specifically, we use a {\em word2vec} vector
model~\cite{mikolov2013efficient} pre-trained on parts of the Google
News dataset~\cite{embedding-google} and calculate the cosine distance
between the \textit{word2vec} representation of the given word and
other words in our chart database.
For example,
 \vspace{-0.05in}
\begin{verbatim}
{"data":{"field": {"wordSim": "population"}}}
\end{verbatim}
 \vspace{-0.05in}
computes the  distance between the \textit{word2vec}
representation of data field name ``population'' and each of the data
field names in our corpus of visualizations.  We consider the data
field to be relevant whenever the cosine similarity between them is
greater than a threshold $\tau$, which we empirically set to $0.75$.
As shown in Supplemental Materials: Figure 3, the charts resulting from this
query are all related to population but contain data fields named
``population'', ``people'', and ``human''.

\begin{figure*}[t!]
	\centering
	\vspace{-0.04in}	
		\includegraphics[width=.98\linewidth]{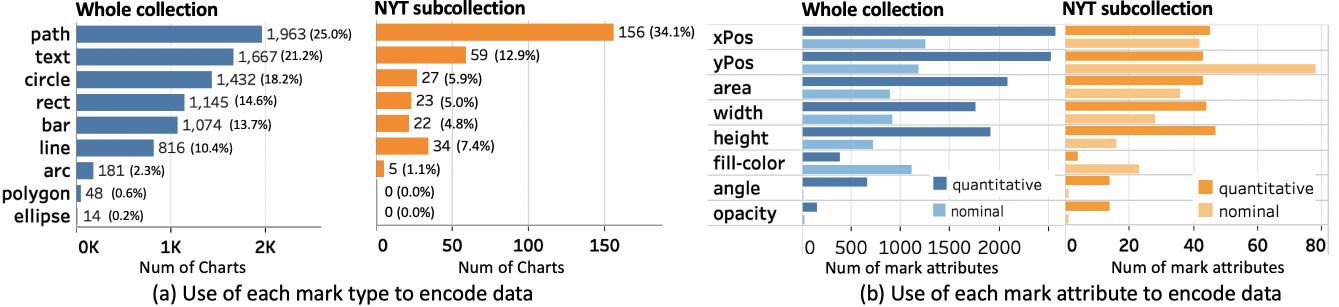}
	\vspace{-0.15in}
	\caption{Number of charts using each mark type for encoding data (a). The percentages are with respect to the total number of charts in each collection (7860 whole collection, 457 New York Times subcollection). Note that some charts contain multiple data encoding marks while others do not contain any encodings. Number of times each mark attribute is used to encode quantitative or nominal data (b).}
	\vspace{-0.2in}
	\label{fig:mark_num_attr_num}
\end{figure*}


\vspace{-0.08in}
\section{Design Demographics Analysis}

We have used our search engine to perform a large scale demographic
analysis across our collection of 7860 D3 visualizations. For
comparison, we independently analyze the subcollection of 457 D3
charts we crawled from the New York Times.  Our goals in conducting
these analyses is to identify patterns of visual design that are
commonly used by D3 chart developers and to show how other researchers
might use our search engine to examine visualization collections at
scale.
Our analysis suggests that visualization developers often adhere to
best design practices as they create D3 visualizations, even if they
only implicitly know these practices.



\vspace{0.04in}
\noindent \textbf{Use of marks:} We analyze the number of charts using each mark
types to encode data (Figure~\ref{fig:mark_num_attr_num}a).
Across our whole collection of D3 charts we find that {\em path} marks
are most frequent (25.0\%), as they are commonly used to generate line
charts and the regional boundary shapes (e.g. state boundaries) used
in choropleth maps. {\em Text} marks (21.2\%) are used to show the
data in text form. {\em Circle} marks (18.2\%) are commonly used in
scatterplots. {\em Rect} marks (14.6\%) are used for heatmaps,
treemaps, and legends, while {\em bar} marks (13.7\%) appear
exclusively in bar charts.  {\em Line} marks (10.4\%),
{\em arc} marks (2.3\%) as used in donut and pie charts,
{\em  polygon} marks (0.6\%) and {\em ellipse} marks (0.2\%) are used to
encode data far less frequently.
Our mark type frequencies are consistent with Battle et
al.'s~\cite{Begel-Baettle-2017} chart type analysis of 1247 D3 charts,
who find that map charts (30.4\%), line charts (12.6\%), bar charts
(12.3\%) and scatterplots (9.46\%) are the four most common types of
D3 charts while pie charts, donut charts and radial charts appear
infrequently (together $<$ 0.05\%).

In our New York Times subcollection
(Figure~\ref{fig:mark_num_attr_num}a right) we find a different
pattern of mark usage; they use {\em paths} (34.1\%) far more often
than the other marks and use far fewer {\em circles} (5.9\%), {\em
  rects} (5.0\%) and {\em bars} (4.8\%).  Manually inspecting the New
York Times subcollection, we find that it contains fewer scatterplots,
heatmaps, treemaps, and bar charts than the collection as a whole, but
often includes more elaborate, and unconventional charts such as
maps, networks and trees which typically involve lots of {\em paths}.

\vspace{0.04in}
\noindent \textbf{Use of mark attributes:} We also analyze the
frequency with which different mark attributes are used to encode
quantitative and nominal data (Figure~\ref{fig:mark_num_attr_num}b).
Across both collections (Whole and New York Times) we find that the pattern of
attribute use to encode quantitative data
is similar and similar to the perceptual effectiveness rankings of visual
attributes proposed by
Mackinlay~\cite{mackinlay1986automating}.  He suggests that
{\em position} is the most effective encoding for quantitative data,
followed by {\em width} and {\em height}, and then {\em angle}, {\em
  area}, {\em fill-color}, and {\em opacity}.
Although we find more {\em area} encodings in our collections than
either {\em height} or {\em width} we note that bar charts redundantly
encode data using either height (or width) and area -- because the
non-data-encoding size attribute of the bars (either {\em width} or
{\em height}) is fixed making {\em area} proportional to the
data-encoding attribute.  If we combine the usage of {\em height} and
{\em width} encodings we find they are used more often than {\em area}
encodings and in-line with Mackinlay's rankings.  These findings
suggest that D3 developers may be aware (perhaps implicitly) of the
perceptual effectiveness rankings when they are designing charts.

\vspace{0.04in}
\noindent \textbf{Use of multiple encodings:}
Visualization developers sometimes choose to encode multiple data
fields using different attributes of the same mark
(Supplemental Materials: Figure 4a).
Across our whole
collection, we find that while some data-encoding marks (27.8\%) are
used to encode a single data field the majority of such marks (72.2\%)
encode multiple data fields.
Using two or more mark
attributes to encode data is far more frequent for {\em circles} than
for any other mark type, as developers often use {\em fill-color} and
{\em area} in addition to {\em position} to encode data in
scatterplots and bubble charts.  While two or three encodings per mark
is quite prevalent, occurrences of five or more encodings per mark is
rare (less than 1\% of marks in our whole collection), as such designs
can make it difficult to interpret the chart.
The pattern of using multiple encodings is similar in our
New York Times subcollection (Supplemental Materials: Figure 4a right).
	
We also analyze which pairs of visual attribute developers commonly
use when encoding multiple quantitative data fields using the same
mark (Supplemental Materials: Figure 4b).
We find that they primarily
follow the approach of using the most perceptually effective visual
attributes~\cite{mackinlay1986automating, munzner2014visualization}
for encoding quantitative data -- i.e. combinations of {\em x-}, {\em
  y-position}, {\em area}, {\em width} and {\em height} are most
prevalent. Again, the pattern is similar for the New York Times
subcollection (Supplemental Materials: Figure 4b right).

\vspace{0.04in}
\noindent \textbf{Use of integral/separable attribute pairs:}
One challenge with multiple encodings is that one encoding may
interfere with the perceptual effectiveness of another encoding.
Graphical perception researchers
call pairs of attributes {\em integral} if viewers have
difficulty reading one without interference from the other and
call them {\em separable} otherwise~\cite{Ware,Palmer}. To avoid interference
attributes that are integral should ideally encode data that is highly
correlated while attributes that are separable can encode correlated or uncorrelated
data. Supplemental Materials: Figure 4c
shows for each pair of data-encoding attributes, how well the
corresponding data fields are correlated.
We find that the Pearson correlation coefficients cover the whole
range of possible values for the first three pairs of attributes. For
many of the pairs lower in the list, the encoded data fields tend to
either be uncorrelated (near 0 correlation) or highly correlated (near
1 or sometimes -1) with a few scattered examples in between. Pairs of
attributes that are known to be highly integral such as width and
height are primarily used to encode highly correlated data fields.
The pattern is even stronger for our New York Times subcollection
(Supplemental Materials: Figure 4c right).
These results suggest
that visualization developers, especially at the New York Times, typically follow
good practices of using integral pairs of encoding attributes when the
data fields are correlated.

%

\vspace{0.04in}
\noindent \textbf{Use of {\em fill-color} encodings:}
Visualization developers and researchers have established
best practices for using {\em fill-color} to encode data.
For example, a rule of thumb is to limit the
number of colors used to encode nominal data to between six and
twelve, so that viewers can perceptually discriminate between
them~\cite{munzner2014visualization, Ware}. Using our search engine we
find that among the visualizations that use {\em fill-color} to encode
nominal data, 58.0\% use six or more different colors while 13.6\% use
twelve or more different colors (Supplemental Materials: Figure 5). However, in the New York Times
subcollection, we find that there was not a single chart using eight
or more different colors to encode nominal data but that 47.9\% use
six or more different colors.  These results suggest that in practice
developers commonly use 6 or more colors for nominal encodings, but
avoid using 12 or more colors.




\section{User Study}

To better understand the effectiveness of our search engine at helping
developers explore the style and structure of
visualizations, we conducted a user study to compare it with a
baseline interface (Sightline\,\cite{sechler2017sightline}) that allows
keyword search over webpages containing D3 visualizations,
but not over the data, marks, encodings and style attributes.
%


\vspace{-0.02in}
\subsection{Study Design}

We designed a comparative, within-subjects study with twelve
participants (ages 18 to 44, four female). They were data analysts,
professional developers, and students who develop data visualizations
frequently as a part of their work to explore data and present their
findings. All participants were experienced with visualization
development using programmatic tools like D3~\cite{d3js},
Vega-Lite\cite{satyanarayan2017vega}, R/ggplot2~\cite{ggplot}, and
Matplotlib~\cite{matplotlib}.  They also reported that they frequently use chart construction software like Microsoft Excel~\cite{ excel} and Tableau~\cite{Tableau}.

Each participant performed two 
exploratory search tasks, one with our search engine interface and one with the baseline interface (Sightline\,\cite{sechler2017sightline}), in counterbalanced order across participants.  Each task consisted of a scenario in which the participant would work as a visualization developer and needed to find 
five visualizations that matched some design criteria.  For instance, in one task, participants were given a dataset with one quantitative and two nominal data fields and they were asked to look for five different charts that might be suitable for showing that dataset.  They were told that the five results should serve as design inspirations and that they should look through the corresponding code to consider how they might reuse or adapt it to their own design.


Each participant completed a pre-study background questionnaire. We
then gave a brief introduction explaining the capabilities of two
search interfaces. For our search interface, we introduced the query
syntax and explained how to specify search over data, marks,
encodings, axes and non-data-encoding attributes as described in
Section~\ref{sec:interface}. Participants could also access a
tutorial page with a written description of the query syntax at any
time. 
After completing each task, we asked
participants to complete a post-task questionnaire.
The study lasted approximately 45 minutes and each
participant was paid \$15. 

\vspace{-.02in}
\subsection{Results}

\begin{figure}[t!]
	\centering
	\includegraphics[width=1\linewidth]{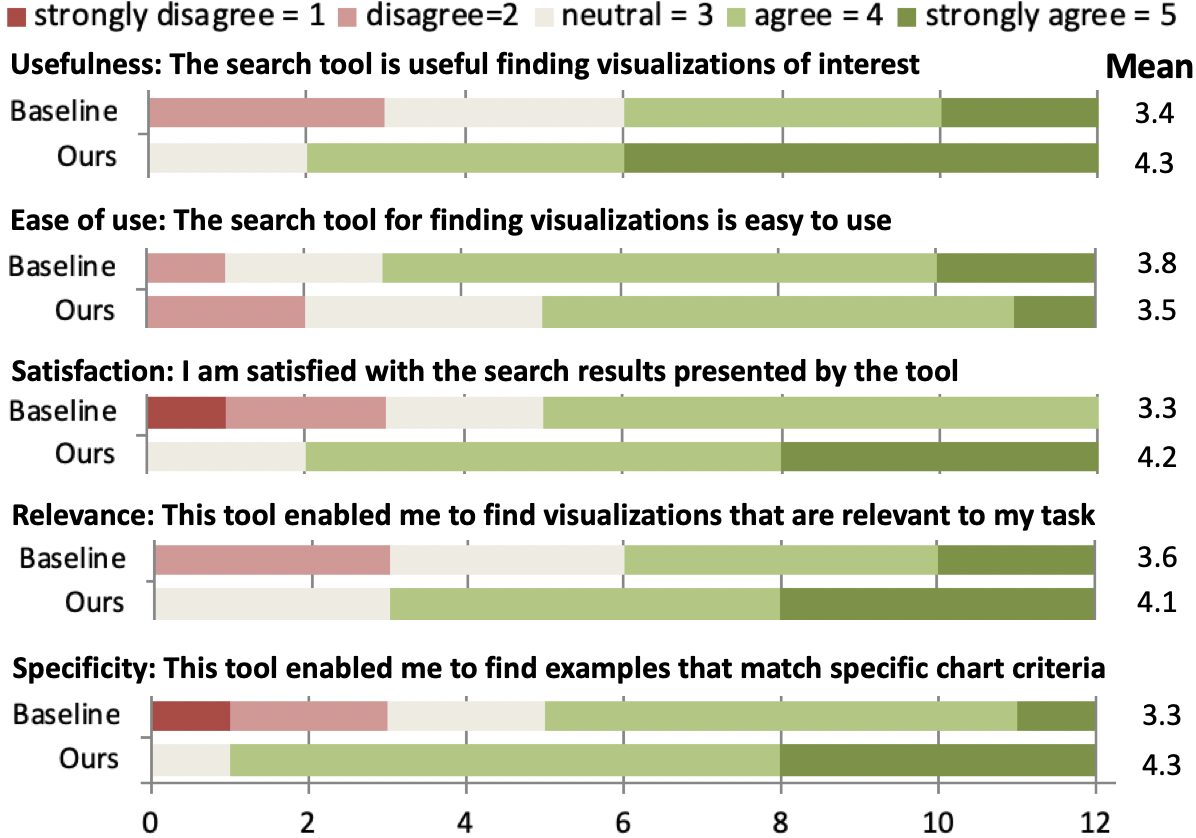}
	\vspace{-0.3in}
	\caption{User study responses to post-task questionnaires.}
	\vspace{-0.22in}
	\label{fig:questionnaire}
\end{figure}

In the post-task questionnaires, participants rated each interface
(baseline and ours) with respect to five measures ({\em Usefulness,
  Ease of use, Satisfaction, Relevance, Specificity}) on 5-point Likert
as shown in Figure~\ref{fig:questionnaire}.  Our interface received significantly higher ratings on three of the five measures:
\textit{Usefulness} (Mann-Whitney $U=37;p=0.035$),
\textit{Satisfaction} ($U=37;p=0.027$) and \textit{Specificity}
($U=38;p=0.031$). For the \textit{Relevance} measure, our interface
received a higher rating overall, but the difference was not
significant ($U=52;p=0.223$). There was no significant difference
between the rating for ease of use ($U=57;p=0.342$).
We also found no significant difference in the task completion time
  for our interface (M = 324 sec., SD = 48) and the baseline (M=311
  sec., SD=41) conditions according to the t-test; t (22)=0.68, p =
  0.501.
Overall in
comparison to the baseline, these results suggest that our
participants found our search interface to be more useful, they were
more satisfied with our search results and that it enabled them to find
examples that match specific chart criteria.
Most participants started with simple queries based on mark type and
one or two encoding criteria. They primarily used the thumbnail view
of the interface for browsing the results and occasionally clicked on
the result items for further examination and to visit the webpage from
where the visualization was crawled. Once they became more familiar
with the query syntax, they sometimes added more criteria
(e.g. non-data encoding attribute such as background) to fulfill their
specific search needs. While some participants had difficulty in
formulating the query (especially at the beginning), gradually they
became comfortable with the help of the auto-complete suggestions
triggered by our search interface as well as by accessing the
tutorial.  Four participants used the query-by-example feature; they
initially gave a text query but then selected a result chart that
looked relevant and submitted it as a query to find other similar
charts.
For the baseline, participants usually started with basic keywords
focused on chart types they thought might be useful for the task
(e.g. `bar', `stacked bar', `scatter plot'),
and then scanned through the paginated list of
results.
But because chart type
does not always appear in the surrounding webpage text or its metadata,
the baseline interface only returned a subset of charts of the given
type. 

In an open-ended feedback session, participants told us that using our tool they were able to find a variety of charts that were
	relevant to the given task.  Overall, they felt that our search engine enabled them to find charts based on design criteria and this would save them hours of time in chart design and
	implementation. One participant mentioned {\em ``I sometimes spend nearly a whole day to find ideal example visualizations to use their code for my visualization development task, but this tool can
		exactly solve that problem within minutes.''} Another participant said, {\em``I am satisfied in finding so many cool visualizations on
		the Web and now I can really use them in my day to day work.''}


The participants also identified a couple of areas of improvement for
our interface.  A few participants felt that they needed time to get
familiar with the query syntax and that coming up with queries that
represented their complex design goals was sometimes challenging. One
participant suggested providing a more diverse gallery of query
examples in the tutorial page to provide further help in formulating
the query and another participant suggested to introduce faceted
search techniques for filtering search results more easily.


\vspace{-.02in}
\section{Limitations}
While our search engine for D3 visualizations demonstrates a broad
range of applications and enables large-scale demographic analysis, it
does have some limitations.


\vspace{0.03in}
\noindent {\bf \em Limitations of D3 deconstructor.}
The deconstruction tool we have extended from Harper and
Agrawala~\cite{harper2014deconstructing, harper2017converting} can
accurately extract the data, marks and encodings  
from a variety of basic D3 charts including variants of bar charts,
scatterplots, line charts, etc. However, it cannot fully
recover certain types of complex encodings such as non-linear function
mappings (e.g., logarithmic scale) between data and mark attributes.
In other cases, charts may use complex
algorithmic techniques for encoding the data (e.g., force-directed layout
algorithm) and their deconstructor cannot recover such mappings.
Our deconstructor inherits these limitations. But despite these issues, 
our search engine does enable exploration over a broad range of the
visualization design space for the most commonly used charts and graphs.

\noindent {\bf \em Limitations of indexing and retrieval.}
Our system indexes D3 visualizations based on their data, marks, and
encodings, axes, non-data-encoding attributes as well as the page text
surrounding the visualizations. However, we do not currently index the
JavaScript code that constructs the visualization. Automatic analysis
of D3 code could provide additional ways of indexing and querying the
visualizations. For example, developers might benefit from finding
pieces of code where a particular D3 function appears in order to
learn how to use it in different contexts. 
This approach may also allow our search engine to index and retrieve visualizations that exhibit a particular interactive technique and/or animation.
Recent work on finding API
usage in programming code~\cite{glassman2018visualizing} could provide
a starting point along this direction.

\vspace{0.02in}
\noindent {\bf \em Limitations of query syntax.}
While our query syntax is designed for visualization developers who are familiar with
  mapping-based specifications (e.g. Vega-Lite), novice users may have difficulties expressing their design
  needs using our current query syntax. Introducing natural language queries
  could make it easier for such users to find visualizations. For example,
  given a query (e.g. ``vertical bar charts'') the
  system might automatically generate the corresponding Vega-Lite
  specification for it and then apply the query to retrieve the corresponding visualizations. Seeing the query in the Vega-Lite syntax could help the user  understand the conversion and allow for further refinement of the query as necessary.  However, converting the natural language query into our
  query syntax could also introduce new challenges (e.g. ambiguity
  in natural language could open up multiple possible interpretations). Another
  direction for supporting novice users is to enhance our
  query-by-example feature, where the interface might allow users to
  select a template visualization and then modify the template by
  changing styles and visual encodings, as demonstrated by Saket et
  al.~\cite{Saket-vis-demonstration}, to find visualizations that match
  their desired template.

\vspace{0.02in}
\noindent 
{\bf \em Limitations of user study design.}
Due to the
  exploratory nature of search tasks, we focused our user study on
  collecting and analyzing subjective data.  While such subjective data 
  does provides information about perceived usefulness, 
  a set of objective metrics
  such as time to construct the query, time to complete the search,
  the variety of charts found and accuracy (e.g. precision) could
  complement the results and is an important direction for future work. Longitudinal studies over an extended period
  of time would also help us further understand the usefulness and adoption
  rate among real users in the future.

\vspace{-.03in}
\section{Conclusions and Future work}
\vspace{-.02in}
We have presented a search engine that indexes and retrieves D3
visualizations based on the data, marks, encodings, axes,
non-data-encoding attributes and webpage metadata.  We demonstrate
that the resulting search engine enables visualization developers
to explore the visualization design space and find visualizations that
match certain data and design characteristics. Users can query our
collection of D3 visualizations in a variety of new ways, encoding
based search, to query by styles to query by example. We believe that
this work is an initial step towards searching and analyzing
visualizations at-scale based on their data and their visual
encodings.  Our search engine system is available at https://www.yorku.ca/enamulh/vis-search.  

There are several avenues for future work. 
Extending the deconstructor to handle
other SVG-based visualizations such as those generated by Plotly~\cite{plotly}, Chartblocks~\cite{chartblocks} and Graphiq~\cite{graphiq}, or indexing the collection of charts used in Beagle~\cite{Begel-Baettle-2017}  would
further enrich our collection and enhance the generalizability of our search engine. The number of Vega-Lite~\cite{satyanarayan2017vega}  charts is growing rapidly and since our query syntax is based on this language already we plan to index these charts in the immediate future.
There are also several collections of image-based charts available in the research community such as MassVis~\cite{borkin2013makes}, VizioMetrics~\cite{lee2016viziometrix} and FigureSeer~\cite{siegel2016figureseer}. Applying recent methods for deconstructing image-based charts would allow us to index and search through such large collection of visualizations~\cite{poco2018extracting, poco2017reverse, jung2017chartsense}.
Moreover, adding different diverse collections could help us to discover differences in design demographics patterns across a wider variety of sources.

\vspace{-.04in}
\acknowledgments{
\vspace{-.03in}	
The authors thank Dae Hyun Kim for feedback over the
course of this work.  This work was supported by NSF award
III-1714647.
}
\bibliographystyle{abbrv-doi}

\bibliography{template}


\section*{Supplementary material}

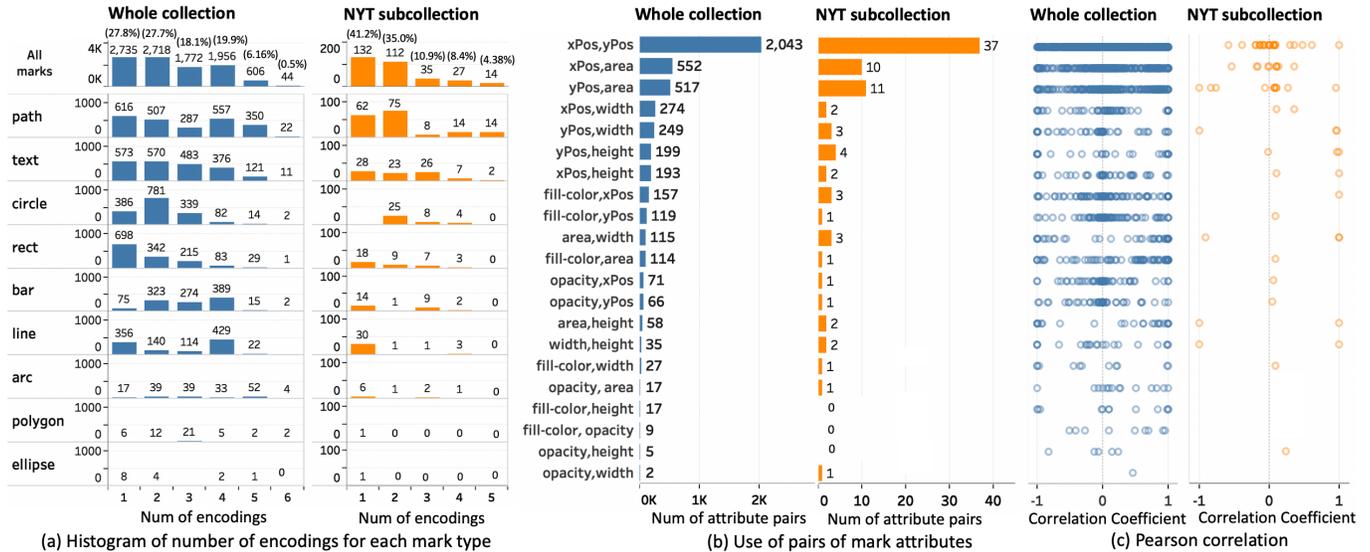

Fig. 4. Histogram of number of encodings for each mark type (a) showing that while 27.8% of data-encoding marks encodes a single attribute, 72.2% encode more than one attribute. Across the whole collection the maximum number of encoding attributes for a mark is 6, whereas for the New York Times subcollection it is 5. Counts of the numbers of data-encoding attribute pairs for the same mark (b) showing that position, area, height and width are most commonly used together. For each data-encoding attribute pair the correlation coefficients for the underlying data (c) show that pairs of attributes that are integral tend to depict correlated (coefficient is 1 or -1) data.

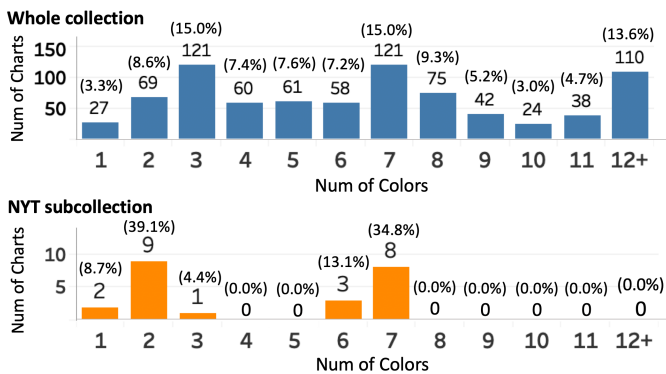

Fig. 5. Histograms of the number of colors used when when fill-color is used to encode nominal data.

subtree of the DOM from the root SVG node for the visualization to the children and look up the non-data-encoding properties of each such node using an SVG parser library [6].

### 4.3 Extracting axes and their attributes

Harper and Agrawala's deconstructor identifies axes by assuming that D3 groups the marks of an axis and stores a *axis scale object* with the group. Their algorithm checks whether each SVG group node has a scale object and if so all the marks under that group are considered as parts of an axis (e.g. axis line, tick marks, tick labels).

In D3 Versions 4 and 5 (V4 and V5) axes are created and the corresponding *scale objects* are stored in a different way. For example, in V3 the developer had to use two separate constructors namely `d3.svg.axis()` and `axis.orient()` to initiate the axis. In contrast, in later versions, the developer just needs to provide one of the four constructors for each orientation: `d3.axisTop()`, `d3.axisRight()`, `d3.axisBottom()`, `d3.axisLeft()`. Also, V4 changes how it stores the *scale object* associated with each axis. With V3, the *scale object* is accessible through `__chart__`. In V4 and V5, the *scale object* is stored as `__axis__` instead. We updated our deconstructor so that if it does not find `__chart__`, it checks if `__axis__` exists and if so it checks the axis orientation ``orient''. If the orientation is either *top* or *bottom*, it considers the axis to be an x-axis, and otherwise (either *left* or *right*) it is considered a y-axis.

Harper and Agrawala's deconstructor also does not identify axis titles which are used to describe the data represented by an axis (e.g. `"title": "date"` for x-axis and `"title": "average price"` for y-axis). In order to extract such axis titles we first identify the set of SVG text elements that do not encode data. For each SVG text element that our deconstructor has marked as not encoding data we use a rule-based approach to check whether it represents a title for the x- or y-axis. We compute the proximity of the candidate text element to the axis tick marks and axis lines. Specifically, if the distance between the y-coordinate of the candidate text and y-coordinate of the x-axis tick marks is under a threshold $t$ and the x-coordinate of starting point of the text element is within the boundary of the x-axis line we consider the text to be the x-axis title. We perform the analogous check for proximity the y-axis. We empirically set the value of $t = 50$. Any text elements that do not meet either of these two criteria are labeled as 'other text labels' (e.g. chart title, chart annotation text).

## 5 USER STUDY: TASKS

**Task 1:** You are a visualization developer and you have a dataset containing one quantitative and two categorical data fields. You are interested in creating a chart that shows this dataset. For this purpose, you need to find a set of example charts that would provide ideas for your design and whose D3 cdoe you might reuse or adapt for your purpose. Your task is to find five different charts that are suitable for showing your dataset using the given search interface.

**Task 2:** You are a visualization developer and you have a dataset containing at least two quantitative and one categorical data fields. You also have over 100 data points in your dataset. You are interested in creating a scatterplot that shows the dataset. For this purpose, you need to find a set of example charts that would provide ideas for your design and whose D3 cdoe you might reuse or adapt for your purpose. Your task is to find five different charts that are suitable for showing your dataset using the given search interface.



\end{document}